%% file: main.tex
\documentclass[conference]{IEEEtran}
\usepackage{amsmath,amsfonts}
\usepackage{booktabs}
\usepackage{algorithmic}
\usepackage{graphicx}
\usepackage{textcomp}
\usepackage{xcolor}
\usepackage{url}
\usepackage{adjustbox}
\usepackage{multirow}
\usepackage{enumitem}
\usepackage[linesnumbered,ruled]{algorithm2e}
\usepackage{bm}
\usepackage{setspace}

\usepackage{flushend}
\usepackage{balance}

\newlength\mylen

\usepackage[caption=false]{subfig}

\def\BibTeX{{\rm B\kern-.05em{\sc i\kern-.025em b}\kern-.08em
		T\kern-.1667em\lower.7ex\hbox{E}\kern-.125emX}}

\newcommand{\lee}[1]{\textcolor{black}{#1}}

\begin{document}

\title{Accelerating Sparse Matrix-Matrix Multiplication on GPUs with Processing Near HBMs}

\author{Shiju Li\textsuperscript{*}, Younghoon Min\textsuperscript{*}, Hane Yie\textsuperscript{*}, Hoshik Kim\textsuperscript{*}, Soohong Ahn\textsuperscript{§}, Joonseop Sim\textsuperscript{§}, Chul-Ho Lee\textsuperscript{‡}, Jongryool Kim\textsuperscript{*} \\
\textsuperscript{*}SOLAB, SK hynix America, San Jose, USA \\
\textsuperscript{§}AMS, SK hynix, Icheon, Korea\\
\textsuperscript{‡}Texas State University, San Marcos, USA \\
{shiju.li, younghoon.min, stella.yie, hoshik1.kim, soohong.ahn, joonseop.sim, jongryool.kim}@sk.com \\
{chulho.lee}@txstate.edu}
\maketitle



\begin{abstract}
Sparse General Matrix-Matrix Multiplication (SpGEMM) is a fundamental operation in numerous scientific computing and data analytics applications, often bottlenecked by irregular memory access patterns. This paper presents Hash-based Multi-phase SpGEMM on GPU and the Acceleration of Indirect Memory Access (AIA) technique, a novel custom near-memory processing approach to optimizing SpGEMM on GPU HBM. Our hardware-software co-designed framework for SpGEMM demonstrates significant performance improvements over state-of-the-art methods, particularly in handling complex, application-specific workloads. 

We evaluate our approach on various graph workloads, including graph contraction, Markov clustering, and Graph Neural Networks (GNNs), showcasing its practical applicability. For graph analytics applications, AIA demonstrates up to 17.3\% time reduction from the software-only implementation, while achieving time reduction of 76.5\% for Graph Contraction and 58.4\% for Markov Clustering compared to cuSPARSE. For GNN training applications with structured global pruning, our hybrid approach delivers an average of 1.43$\times$ speedup over software-only implementation across six benchmark datasets and three architectures (GCN, GIN, GraphSAGE), and shows 1.95$\times$ speedup for GNN workloads when compared to cuSPARSE, with up to 4.18$\times$ gains on large-scale datasets. 

\end{abstract}
\maketitle



\input{sections/intro}
\input{sections/related}
\input{sections/SpGeMM_on_GPU}
\input{sections/system}

\input{sections/algorithm}
\input{sections/result}
\input{sections/conclusion}
\bibliographystyle{IEEEtran}
\bibliography{ref}
	
\end{document}

%% file: sections/intro.tex
\section{Introduction}

Sparse General Matrix-Matrix Multiplication (SpGEMM) plays a pivotal role in computational science, machine learning, and graph \lee{analytics}. \lee{The general matrix-matrix multiplication (GEMM), which multiplies a $m \times k$ matrix $\mathbf{A}$ with a $k \times n$ matrix $\mathbf{B}$ to produce a $m \times n$ matrix $\mathbf{C} = \mathbf{A}\mathbf{B}$, is a simple yet fundamental matrix operation in many applications such as algebraic multigrid methods~\cite{bell2012exposing}, multi-source breadth-first search~\cite{bulucc2011combinatorial}, recursive formulations of all-pairs shortest-paths algorithms\cite{d2007r}, clustering coefficient~\cite{azad2015parallel}, graph contraction~\cite{gilbert2008unified}, Markov clustering~\cite{van2000graph}, and graph neural networks (GNNs)~\cite{ma2021deep,hamilton2020graph,huang2022characterizing}. However, these applications often require exploiting \emph{sparsity} in both input and output matrices to optimize storage and computational efficiency. GNNs are a compelling application for SpGEMM optimization because they rely on sparse matrix operations for feature aggregation and propagation across neighborhoods. Recently, it has been shown in~\cite{peng2024maxk} that optimized sparse operations can achieve an average of 3.2$\times$ speedup over state-of-the-art GNN frameworks.}


However, the efficient execution of SpGEMM on modern GPUs faces several fundamental challenges, despite GPUs offering superior peak floating-point performance and memory bandwidth compared to CPUs. \lee{The primary challenges arise from the irregular nature of sparse computations. The irregular memory access patterns create a strong mismatch between data access patterns and memory layout, leading to limited spatial locality and cache effectiveness.} While compressed representations of sparse matrices reduce memory footprint, they incur significant overheads in accessing and processing metadata. \lee{These challenges are manifested in three critical aspects: (1) the unknown number of non-zero entries in the resulting matrix (or matrix product) prior to computation, (2) expensive parallel insert operations at random positions within the resulting matrix, and (3) load balancing issues that arise when handling input matrices with diverse sparsity structures.}


Despite attempts to address these challenges with various approaches, including dedicated hardware accelerators and software optimizations~\cite{naumov2010cusparse,dalton2015optimizing}, previous GPU SpGEMM methods have shown limited success. They either perform well only for relatively regular sparse matrices or introduce significant memory overhead for specific sparsity patterns~\cite{demouth2012sparse,bell2012exposing}. Moreover, existing GPU implementations have struggled to consistently outperform well-optimized CPU approaches, highlighting the difficulty in fully utilizing GPUs' massive parallelism for sparse matrix operations~\cite{wang2014intel}.

\lee{On the other hand, the advancement in 3D integration technologies has made it more viable to couple compute units close to the memory, a concept known as processing near memory (PNM)~\cite{singh2019near}. Processing right at the ``home'' of data can significantly reduce data movement. Furthermore, modern GPUs are equipped with high bandwidth memory (HBM)~\cite{lee201425}, which adopts a vertically stacked architecture that allows for greater capacity and bandwidth. Therefore, we leverage the concept of processing near GPU HBM to effectively tackle the challenges of SpGEMM.}


In this paper, we propose and implement the Acceleration of Indirect Memory Access (AIA) solution, a novel \lee{processing-near-HBM approach} to enhance performance in handling indirect memory access patterns in SpGEMM operations\lee{, thereby accelerating the overall computation}. We introduce the AIA technique, designed to assist the primary computational core in sparse data computations, specifically targeting SpGEMM operations on HBM GPUs. Our work makes three key contributions through the co-design of hardware and software for accelerating SpGEMM on GPUs:


\begin{itemize}[itemsep=2pt,leftmargin=1.5em,topsep=3pt]
\item We implement an optimized Hash-Based multiphase\\       SpGEMM algorithm on H200 GPU that systematically addresses the fundamental challenges of sparse matrix multiplication through intelligent workload distribution and adaptive memory management. Our three-phase approach (row-grouping, allocation, and accumulation) employs sophisticated hash tables with collision resolution and dual thread assignment strategies (PWPR and TBPR) optimized for different workload characteristics. 

\item We introduce a novel Processing-Near-HBM hardware technique, Acceleration of Indirect memory Access (AIA), that transforms SpGEMM's irregular memory access patterns into efficient sequential streams. The AIA solution, integrated between GPU cores and HBM stacks, implements ranged indirect access functionality specifically designed for SpGEMM's two-level indirection patterns. Our hardware solution improves L1 cache hit ratios from 64.41\% to 75.14\% in accumulation phases and from 64.66\% to 88.15\% in allocation phases.

\item Our hybrid hardware-software solution delivers substantial improvements across diverse workloads. For matrix self-product operations, our AIA-enhanced implementation achieves an average 80.5\% runtime reduction and 6.87× throughput improvement compared to cuSPARSE. For graph analytics applications, AIA demonstrates an average of 76.5\% improvement for Graph Contraction and 58.4\% for Markov Clustering compared to cuSPARSE. Most significantly, for GNN training with structured pruning, our approach delivers 30.3\% average training time reduction over the software-only implementation, and 48.6\% over cusparse, across diverse GNN benchmark datasets and three GNN model(GCN, GIN, GraphSAGE), with large-scale datasets like Products achieving up to 76.1\% speedup. These results establish the practical impact of our co-designed solution for production-scale applications.

\end{itemize}

%% file: sections/related.tex
\section{Background}
\subsubsection*{Memory Wall and Indirect Memory Access.} 
Memory wall represents a critical bottleneck in modern computing systems, particularly affecting graph analysis, sparse linear algebra, and various other applications~\cite{chen2018tvm,kocberber2013meet,mislove2007measurement}. This challenge primarily stems from irregular memory access patterns, where data are arbitrarily accessed with poor locality, resulting in frequent cache misses and high latency when fetching data from the memory hierarchy. A key pattern contributing to this bottleneck is indirect memory access, typically appearing as $x[a[i]]$, where array $a$'s values serve as indices to access array $x$. This pattern becomes even more complex in the form of ranged indirect access $(x[a[i]], x[a[i]+1], x[a[i]+2], ...)$, commonly seen in graph algorithms processing adjacency lists, where $a[i]$ might point to the start of a vertex's neighbor list in array $x$, and multiple consecutive elements are accessed from that point. Such patterns pose significant challenges for conventional hardware prefetchers, as they require understanding both irregular access patterns and variable-length ranges of accessed data.

\subsubsection*{GPU-Centric Near Memory Processing and Challenges.} 
The integration of near-memory processing with GPU HBM architectures presents both opportunities and significant challenges. Several notable architectures have explored this concept in GPU contexts, with TOM~\cite{hsieh2016transparent} pioneering a novel approach using lightweight GPU cores integrated into 3D-stacked memories, supported by an automatic compiler framework for offload decisions. Building on this foundation, Pattnaik et al.~\cite{pattnaik2016scheduling} advanced the concept with an NMC-assisted GPU architecture featuring an affinity prediction model for kernel execution placement. Recent research has particularly focused on leveraging HBM's unique capabilities for near-memory processing. DNN-PIM~\cite{liu2018processing} introduced a heterogeneous architecture incorporating both programmable ARM cores and fixed-function units in the HBM logic layer. Complementing this work, Boroumand et al. ~\cite{boroumand2018google} evaluated PNM architectures for consumer applications, carefully considering the practical constraints of area and power budgets in HBM implementations.

Despite these advances, four fundamental challenges must be addressed for effective implementation. First, the programming model must efficiently integrate with existing GPU frameworks while providing mechanisms to determine optimal code placement between host and near-memory units ~\cite{liu2018processing,boroumand2018google}. Second, memory management requires sophisticated data mapping strategies between GPU cores and HBM stacks, as demonstrated by MONDRIAN~\cite{drumond2017mondrian}, which emphasized the importance of hardware-software co-design in optimizing memory access patterns. Third, virtual memory support must align with GPU memory management systems while maintaining efficient address translation mechanisms, a challenge effectively addressed in TESSERACT~\cite{ahn2015scalable} and subsequent works~\cite{ahn2015pim}. Finally, cache coherency remains a critical challenge in maintaining consistency between GPU cores and near-memory processing units, with recent work by Farmahini et al.~\cite{farmahini2015nda} proposing innovative solutions for optimizing HBM bandwidth utilization and minimizing unnecessary off-chip traffic.


\subsubsection*{Prior SpGEMM Algorithms.} 
The evolution of SpGEMM algorithms reflects ongoing efforts to optimize sparse matrix multiplication across different hardware architectures. Traditional approaches can be categorized into three main strategies: outer product, inner product, and row-wise methods. The outer product method, exemplified by Dalton et al. ~\cite{dalton2015optimizing}, computes the contribution of each non-zero element in matrix A and B independently, requiring efficient merging of intermediate results. Inner product methods, as implemented by Demouth~\cite{demouth2012sparse}, compute each element of the output matrix through dot products of corresponding rows and columns, facing challenges with memory locality. Row-wise methods, adopted by Naumov et al.~\cite{naumov2010cusparse}, process each row of the output matrix independently, offering better parallelism but requiring careful load balancing strategies.

On GPU platforms, SpGEMM implementations have evolved from basic CSR-based approaches to sophisticated hybrid formats and specialized algorithms. Early implementations struggled with the irregular nature of sparse matrix operations, but recent advances have introduced various optimizations. Hash-based approaches, as demonstrated by Liu and Vinter~\cite{liu2014efficient}, use hash tables to accumulate partial products efficiently. Sort-based methods, exemplified by Nagasaka et al.~\cite{nagasaka2017high}, offer better memory coalescing but incur sorting overhead. Hybrid methods, such as those proposed by Gremse et al.~\cite{gremse2015gpu}, combine multiple strategies to balance computation and memory access efficiency.

Despite these advancements, existing implementations still face fundamental challenges in fully utilizing GPU capabilities for SpGEMM operations. The irregular memory access patterns create a strong mismatch between data access patterns and memory layout, leading to limited spatial locality and cache effectiveness, which causes frequent data movement between memory hierarchies. Processing-Near-HBM solutions offer a promising approach to address these challenges by reducing data movement and enhancing memory access efficiency.

\subsubsection*{SpGEMM Accelerates GNN} 
Recent works have explored several promising approaches to GNNs through SpGEMM. SpGEMM accelerates GNN through strategic transformation of computational patterns and exploitation of induced sparsity. MaxK-GNN~\cite{peng2024maxk} demonstrates the most dramatic approach, converting traditional SpMM operations into SpGEMM by introducing MaxK nonlinearity that sparsifies feature matrices to 87.5\% sparsity. This transformation reduces global memory traffic by over 90\% and achieves $3-7\times$ kernel-level speedups on NVIDIA GPUs, with L1 cache hit rates improving from 1.53\% to 22.16\%. Shivdikar et al.~\cite{shivdikar2024neurachip} also propose a novel GNN spatial accelerator using spgemm on the sparsified feature matrix. PruneGNN~\cite{gurevin2024prunegnn} takes a complementary approach through structured dimension-wise pruning, selectively using SpGEMM for backpropagation steps where both weight and gradient matrices are sparse, achieving $2\times$ average speedup on A100 GPUs through SIMD-aware kernel design. Tripathy et al.~\cite{tripathy2024distributed} reframes GNN neighborhood sampling as a matrix-based bulk sampling technique, expressing sampling as SpGEMM operations to efficiently process multiple minibatches simultaneously. This matrix-based approach allows complex sampling procedures to be represented through a series of SpGEMM operations: computing probabilities, performing rejection sampling, and executing row and column extractions, enabling $2.5-8.46\times$ speedups for distributed training by processing multiple minibatches simultaneously. These implementations address SpGEMM's fundamental challenges—irregular memory access patterns, load imbalance, and accumulation bottlenecks—through techniques including warp-level edge partitioning, shared memory buffering, and row-wise product algorithms that reduce computational complexity from $O(Ed)$ to $O(Ek)$ where $k \ll d$, which collectively demonstrate SpGEMM's critical role in addressing the computational bottlenecks in GNNs, especially for large-scale graphs, while maintaining comparable accuracy to dense implementations.

%% file: sections/SpGeMM_on_GPU.tex
\section{Optimized Hash-based Multi-phase SpGEMM on GPU}

This section presents our optimized SpGEMM algorithm designed to leverage GPU architectural features while addressing the fundamental challenges of sparse matrix multiplication. Our approach employs a multi-phase strategy with intelligent workload distribution and adaptive memory management to maximize GPU utilization.
\subsection{Algorithm Overview}

\lee{Our SpGEMM implementation follows the row-wise product method known as Gustavson’s algorithm~\cite{gustavson1978two} and is based on state-of-the-art hash-based SpGEMM libraries~\cite{nagasaka2017high,parger2020speck,du2022opsparse}. For input matrices $\mathbf{A}$ and $\mathbf{B}$, the algorithm computes the output matrix $\mathbf{C}$ one row at a time. For each row $i$ of $\mathbf{A}$, every non-zero element $A_{i,k}$ is used to scale the entire $k$-th row of matrix $\mathbf{B}$. The row index $k$ of matrix $\mathbf{B}$ is determined by the column index of the non-zero value $A_{i,k}$ from matrix $\mathbf{A}$. All of these resulting scaled rows are then accumulated (added together, column-by-column) into the $i$-th row of the output matrix $\mathbf{C}$.}


\lee{In the implementation, we use a three-phase approach to systematically address the challenges of unknown output sparsity, irregular memory access patterns, and load imbalance. It begins with a row-grouping phase for workload analysis and preprocessing, which is then followed by an allocation phase to learn the structure of the output matrix and allocate GPU resources accordingly. It concludes with an accumulation phase to generate the out matrix. Each phase is carefully designed to exploit GPU parallelism while minimizing memory overheads.}



\lee{Computing $\mathbf{C} = \mathbf{A}\mathbf{B}$ for sparse matrices $\mathbf{A}$ and $\mathbf{B}$ requires processing varying numbers of intermediate products for different output rows. For each row of $\mathbf{C}$, the number of intermediate products, denoted as IP, can be obtained using Algorithm~\ref{alg:intermediate}.} This metric indicates the amount of the workload needed to obtain each row of $\mathbf{C}$, so it is used as a basis for our load-balancing strategy.

\begin{algorithm}[t]
\caption{Intermediate Product Counting}
\label{alg:intermediate}
\KwIn{Matrix $\mathbf{A}$ in CSR format $(rpt_{A}, col_{A})$,
        Matrix $\mathbf{B}$ row pointers $(rpt_{B})$}
\KwOut{$intermediateCount$ for each row}
\For{$i=0$ \KwTo $len(rpt_{A})-1$}{
    $count \gets 0$ \\
    \For{$j = rpt_{A}[i]$ to $rpt_{A}[i+1]-1$}{
        $col \gets col_{A}[j]$\\
        $count \gets count + (rpt_{B}[col+1] - rpt_{B}[col])$
    }
    $intermediateCount[i] \gets count$
}
\end{algorithm}

\vspace{-2mm}
\subsection{Row-grouping Phase}

To address workload imbalance inherent in SpGEMM, we implement a two-stage grouping mechanism that organizes matrix rows based on computational characteristics without physically reordering the matrix.

\vspace{2pt}
\noindent \textbf{Initial Grouping.} The rows of matrix $\mathbf{A}$ are classified into four groups based on the number of intermediate products using logarithmic binning. Then, each row of matrix $\mathbf{A}$ is reordered based on which group it belongs to. We have a map between the original row IDs and their sorted IDs according to their group. Let $Map[i]$ be the original row ID that corresponds to index $i$ in the sorted IDs. 

\begin{table*}[ht]
\centering
\caption{GPU resource allocations for different groups}
\vspace{-2mm}
\begin{tabular}{|c|c|c|c|c|}
\hline
\textbf{Group ID} & \textbf{IP Range} & \textbf{Thread Assignment} & \textbf{Thread Block Size} & \textbf{Hash Table Size} \\
\hline
0 & $0-31$ & PWPR & 512 & 64 \\
\hline
1 & $32-511$ & TBPR & 256 & 1024 \\
\hline
2 & $512-8191$ & TBPR & 1024 & 8192 \\
\hline
3 & $\geq 8192$ & TBPR & 1024 & Global Memory \\
\hline
\end{tabular}
\label{tab:h200-params}
\end{table*}

\begin{algorithm}[t]
\caption{PWPR SpGEMM-Allocation Phase}
\label{alg:pwpr-allocation}
\KwIn{Matrix $\mathbf{A}$: $(rpt_{A}, col_{A})$,
      Matrix $\mathbf{B}$: $(rpt_{B}, col_{B})$,
      $Table[]$,$Map$}
\KwOut{$Table[]$ with unique column indices,
       $uniqueCount$ for row $i$}
$g\_{threadIdx} = blockIdx * blockDim + threadIdx$,\\
$laneIdx \gets threadIdx \% 4$, $i\gets Map[g\_{threadIdx}/4]$\\
\For{$j = rpt_{A}[i] + laneIdx$ \KwTo $rpt_{A}[i+1]-1$ $\mathbf{step}$ $4$}{
    $col \gets col_{A}[j]$\\
    \For{$k = rpt_{B}[col]$ \KwTo $rpt_{B}[col+1]-1$}{
        $key \gets col_{B}[k]$\\
        $uniqueCount\gets$ InsertIntoTable$(key)$
    }
}
\_\_syncwarp()\\
\If{$LaneIdx = 0$}{
    $rpt_{C}[i+1] \gets rpt_{C}[i] + uniqueCount$
}
\end{algorithm}

\begin{algorithm}[t]
\caption{TBPR SpGEMM-Allocation Phase}
\label{alg:tbpr-allocation}
\KwIn{Matrix $\mathbf{A}$: $(rpt_{A}, col_{A})$,
      Matrix $\mathbf{B}$: $(rpt_{B}, col_{B})$, $Table[]$,$Map$}
\KwOut{$Table[]$ with unique column indices,
       $uniqueCount$ for row $i$}
    $warpIdx \gets threadIdx / 32$,~$laneIdx \gets threadIdx \% 32$\\ $\#warps \gets blockDim / 32$,
        $i\gets Map[blockIdx]$\\
\BlankLine
\For{$j = rpt_{A}[i] + warpIdx$ \KwTo $rpt_{A}[i+1]-1$ $\mathbf{step}$ $\#warps$}{
    $col \gets col_{A}[j]$\\
    \For{$k = rpt_{B}[col] + laneIdx$ \KwTo $rpt_{B}[col+1]-1$ $\mathbf{step}$ $32$}{
        $key \gets col_{B}[k]$\\
        $uniqueCount\gets$ InsertIntoTable$(key)$
    }
}

\_\_syncthreads()\\

\If{$threadIdx = 0$}{
    $rpt_{C}[i+1] \gets rpt_{C}[i] + uniqueCount$
}
\end{algorithm}

\vspace{2pt}
\noindent \textbf{GPU Resource Allocation.} Each row in a different group is assigned a different number of threads according to its corresponding workload. The ones with heavier workloads are assigned larger thread blocks. Each group is also processed independently, potentially for better load balancing and resource utilization. Furthermore, a hash table is created for each group in shared memory, where its size is proportional to the group's workload. This hash table is used to obtain the results of intermediate products. Table~\ref{tab:h200-params} summarizes the allocation of GPU resources for different groups.

\subsection{Allocation Phase}
The allocation phase determines the structure of the output matrix $\mathbf{C}$ by identifying unique column indices in each row without computing actual values. Multiple CUDA kernels are launched with different streams in this phase, enabling parallel execution across groups.

\vspace{2pt}
\noindent \textbf{Thread Assignment Strategies.}
We implement two complementary thread assignment approaches optimized for different workload characteristics, namely, partial warp per row (PWPR) and thread block per row (TBPR), as shown in Algorithm~\ref{alg:pwpr-allocation} and Algorithm~\ref{alg:tbpr-allocation}. 

PWPR assigns four threads for each row of matrix $\mathbf{A}$ for the light workload, i.e., Group 0. Each thread processes a portion of non-zero elements from matrix $\mathbf{A}$ and their corresponding rows from matrix $\mathbf{B}$. The outer loop distributes the non-zero elements in matrix $\mathbf{A}$ among the four threads in a cyclic manner (or a strided pattern) for coalesced memory access. Each tread retrieves the non-zero elements at the rows of matrix $\mathbf{B}$, each of which is indicated by the column index of each non-zero element of matrix $\mathbf{A}$. The warp-level synchronization is then done to ensure consistency before updating the row pointers of the output matrix $\mathbf{C}$. 

TBPR assigns each block  (i.e., multiples of warps) for each row of matrix $\mathbf{A}$ for the moderate to heavy workloads, i.e., Groups 1--3. TBPR works in a similar way to PWPR, but based on both warp-level and thread-level parallelisms. At each time, each block processes one non-zero element from matrix $\mathbf{A}$ and its corresponding row from matrix $\mathbf{B}$ in a cyclic fashion. However, unlike PWPR, the inner loop distributes the process of retrieving the non-zero elements at the rows of matrix $\mathbf{B}$, each of which is indicated by the column index of each non-zero element of $\mathbf{A}$, across 32 threads (a warp). This dual-level approach maximizes memory bandwidth utilization--multiple warps process different parts of matrix $\mathbf{A}$ simultaneously while threads within each warp access consecutive non-zero elements of matrix $\mathbf{B}$.

\vspace{2pt}
\noindent \textbf{Hash Table.} Algorithm~\ref{gpu_hash} implements a collision-free hash table using linear probing. It begins with an empty hash table (initialized to $-1$) and a counter to track unique elements. When processing a new column index, it first computes a hash position using multiplication and modulo operations for balanced distribution within the hash table. It then enters a loop that handles three possible scenarios: finding an existing entry, inserting a new entry, and resolving a collision. The hash table employs atomic Compare-And-Swap operations to safely handle concurrent insertions in parallel. 

The size of the hash table is first set to the value of IP obtained from Algorithm~\ref{alg:intermediate} and then determined by the number of columns with nnz entries, i.e., $uniqueCount$. For each row with a large number of nnz entries, we use a two-phase approach. It first attempts to use a shared memory hash table and then falls back to global memory if the shared memory capacity is exceeded. This approach ensures efficient handling of varying row sizes while maximizing performance.

\begin{algorithm}
\caption{InsertIntoTable \& (AddInTable)}\label{gpu_hash}
\KwIn{$key$,  (valA,valB)}
\KwOut{$uniqueCount$,($Table$, $Tableval$)}
$Table[] \gets -1$,$uniqueCount \gets 0, $($Tableval[] \gets 0$)
$hashPos \gets (key * multiplier) \% tableSize$ \\
\While{true}{
    \eIf{$Table[hashPos] = key$}{
        (atomicAdd$(\&Tableval[hashPos], valA \times valB)$)
        \textbf{break}   
    }{\eIf{$Table[hashPos] = -1$}{    
        $oldValue \gets atomicCAS(Table + hashPos, -1, key)$ \\
        \If{$oldValue = -1$}{
            $uniqueCount  \gets uniqueCount + 1$ \\        (atomicAdd$(\&Tableval[hashPos], valA \times valB)$)         
            \textbf{break}
        }
    }{
        $hashPos \gets (hashPos + 1) \% tableSize$
    }
    }
}
\end{algorithm}

\begin{algorithm}[t]
\caption{PWPR SpGEMM-Accumulation Phase}
\label{alg:pwpr-accumulation}
\KwIn{Output From allocation phase, Matrix $\mathbf{A}$: $(rpt_{A}, col_{A}, val_{A})$,
      matrix $\mathbf{B}$: $(rpt_{B}, col_{B}, val_{B}$), $Table[]$,
       $Tableval[]$}
\KwOut{ nnz entry of matrix $\mathbf{C}$ on row $i$:$(col_C,val_C)_i$}
$g\_{threadIdx} = blockIdx * blockDim + threadIdx$,\\
$laneIdx \gets threadIdx \% 4$, $i\gets Map[g\_{threadIdx}/4]$\\

\For{$j = rpt_A[i] + laneIdx$ \KwTo $rpt_A[i+1]-1$ $\mathbf{step}$ $4$}{
    $colIdx_A \gets col_A[j]$\\
    $valA \gets val_A[j]$\\
    \For{$k = rpt_B[colIdx_A]$ \KwTo $rpt_B[colIdx_A+1]-1$}{
        $key \gets col_B[k]$\\
        $valB \gets val_B[k]$\\
        $Table[]$, $Tableval[]$ $\gets$ AddInTable$(key,valA,valB)$
    }
}

\_\_syncwarp()\\

$startPos \gets rpt_C[i]$\\
\For{$idx = laneIdx$ \KwTo $uniqueCount-1$ $\mathbf{step}$ $4$}{
    $col_C[startPos + idx] \gets Table[idx]$\\
    $val_C[startPos + idx] \gets Tableval[idx]$
}
\_\_syncwarp()\\
$(col_C,val_C)_i= BitonicSort($\\$col_C[startPos],...,col_C[startPos+uniqueCount-1],$\\
$val_C[startPos],...,val_C[startPos+uniqueCount-1])$
\end{algorithm}

\subsection{Accumulation Phase}


This final phase combines the results of intermediate products to obtain the elements of the output matrix $\mathbf{C}$ in three steps. Similar to the allocation phase, we also implement two complementary thread assignment approaches, where the PWPR approach is shown in Algorithm~\ref{alg:pwpr-accumulation}. The accumulation algorithm performs the final value computation phase of SpGEMM, building upon the structure determined in the allocation phase. This algorithm maintains the same 4-thread or entire threadblock cooperative approach but now focuses on computing actual numerical results rather than just identifying output structure.

\vspace{2pt}
\noindent \textbf{Value Computation and Accumulation.} The algorithm maintains dual hash tables -- one for column indices and another for accumulated values. As intermediate products are computed, the hash table accumulates contributions to each unique column position. When multiple multiplication paths produce values for the same column index, atomic operations ensure correct accumulation. This phase simultaneously counts the total non-zero elements per row, essential for CSR format construction.

\vspace{2pt}
\noindent \textbf{Element Gathering.} Once accumulation completes, threads collaboratively extract the computed (column, value) pairs from the hash table into contiguous memory. Each thread is assigned specific hash table entries to process, ensuring balanced workload distribution and coalesced memory access patterns during the gathering operation.

\vspace{2pt}
\noindent \textbf{Column Index Sorting.} The final stage sorts the gathered elements by column index to maintain CSR format requirements. Each thread determines its assigned element's position through parallel comparison operations against other elements in the same row. Based on these computed positions, threads write their elements directly to the appropriate locations in global memory.

This three-step organization enables efficient parallel processing while guaranteeing correctness of the CSR output, balancing the competing demands of accumulation accuracy, memory efficiency, and computational throughput.

%% file: sections/system.tex
\section{Near Memory Processing: HBM-AIA}
\begin{figure}[ht]
\centering
\includegraphics[width=0.48\textwidth]{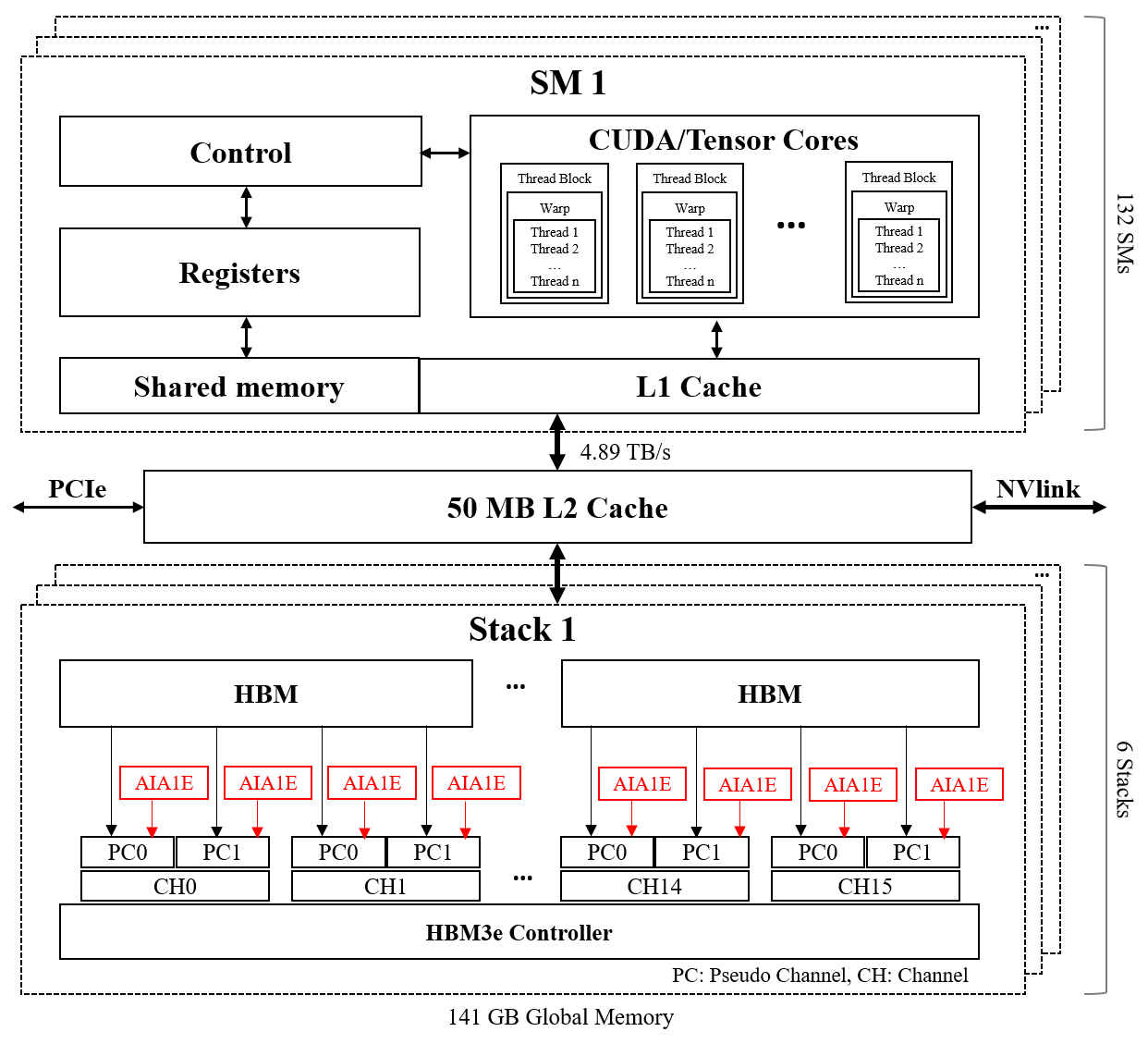}
\caption{H200 GPU with AIA agent on HBM}
\label{fig:system_architecture}
\end{figure}

Our proposed hardware solution integrates Indirect Access capabilities with the GPU's High Bandwidth Memory (HBM) to accelerate Sparse General Matrix-Matrix Multiplication (SpGEMM). This integration leverages the GPU's parallel processing power while addressing the irregular memory access patterns inherent in SpGEMM operations.

\subsection{Motivation of AIA}
Hash-based multi-phase SpGEMM faces three critical bottlenecks. First, the grouping phase requires massive atomic operations on global memory, consuming over 10\% of execution time despite its $O(\#(row))$ complexity. Second, the hashing method's random access patterns cause severe bank conflicts in shared memory. Third, hash table sizing presents an unavoidable trade-off: large tables matching maximum nnz per row cause high collision rates, while uniform sizing underutilizes hardware for smaller rows.

These issues arise from the fundamental mismatch between SpGEMM's irregular memory access patterns and GPU architecture. The AIA technique addresses this by transforming random accesses into sequential streams through processing-near-HBM architecture. By implementing ranged indirect access at the HBM level, AIA converts SpGEMM's two-level indirection into predictable patterns that better utilize the memory hierarchy while reducing data movement and eliminating intermediate memory allocations.

\subsection{HBM-AIA on Modern GPU}
The SpGEMM hardware solution integrates AIA functionality directly into the GPU's memory hierarchy. The figure~\ref{fig:system_architecture} shows the layout of Streaming Multiprocessors (SMs) and HBM Stacks in the H200 GPU. The HBM subsystem consists of 6 stacks (Stack 1 shown in detail), with each stack containing multiple HBM dies. The AIA engines are embedded within the HBM Controller at the base of each stack, positioned strategically between the memory dies and the GPU cores. This placement enables the AIA engines to perform indirect memory access operations directly at the memory level, transforming irregular access patterns into efficient sequential streams before data traverses the memory hierarchy. The 141 GB Global Memory capacity is distributed across the HBM stacks, with the AIA engines managing data gathering operations locally within each stack. This architecture minimizes data movement by processing indirect accesses near the memory source, reducing pressure on the L2 cache and PCIe/NVlink interfaces. The AIA engines work in conjunction with the GPU's cache hierarchy—gathering and coalescing scattered data accesses at the HBM level ensures that when data reaches the L1 and L2 caches, it arrives in more cache-friendly sequential patterns, thereby improving overall cache utilization and reducing memory access latency for sparse matrix operations.


\subsection{AIA Ranged index Illustration}
\begin{figure}[ht]
\centering
\includegraphics[width=0.48\textwidth]{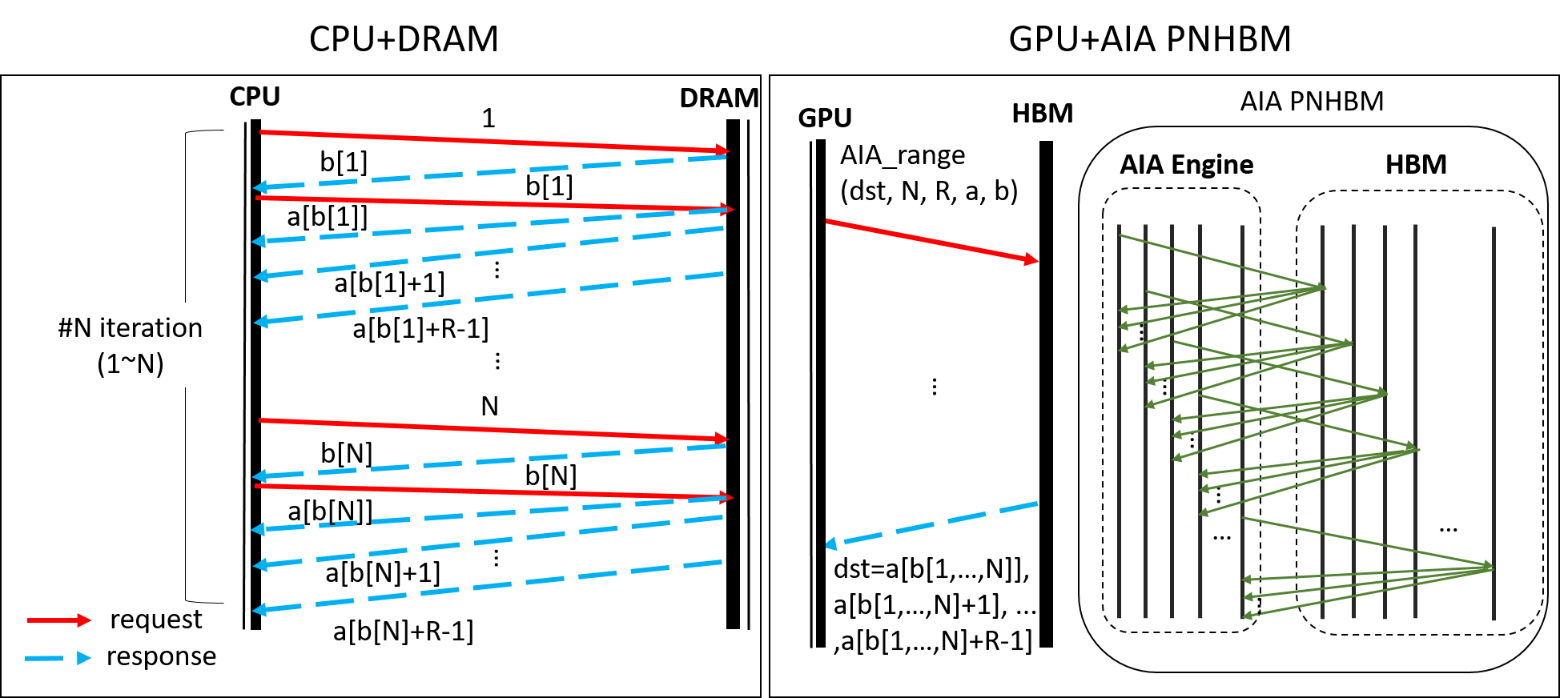}
\caption{AIA ranged index}
\label{fig:AIA_ranged_index}
\end{figure}
Figure~\ref{fig:AIA_ranged_index} compares traditional CPU+DRAM indirect memory access (left) with the GPU+AIA PNHBM approach (right). In the traditional approach, the CPU performs $N$ iterations where each iteration requires two separate memory transactions: first sending a request for index $b[i]$ and receiving its value, then using that value to request the range $a[b[i]]$ through $a[b[i]+R-1]$, resulting in $2N$ total memory round trips with associated latency penalties. In contrast, the GPU+AIA PNHBM approach consolidates this into a single operation where the GPU sends one AIA range request with parameters $(dst, N, R, a, b)$ to the AIA Engine located within the Processing Near HBM (PNHBM), which then performs all $N$ indirect lookups locally near the memory—fetching each $b[i]$ and retrieving the corresponding range $a[b[i]]$ to $a[b[i]+R-1]$ internally—before returning all results in a single bulk response stream. This transformation from 2N processor-memory round-trips to just one request-response pair eliminates the latency overhead of multiple communications, while the AIA Engine's internal switching network (shown in the interconnect diagram) efficiently routes the indirect memory accesses within the HBM stack, converting inherently random access patterns into optimized sequential streams that better utilize memory bandwidth and reduce overall access latency.



\subsection{AIA Kernel on SpGEMM}
\begin{figure}[ht]
\centering
\includegraphics[width=0.48\textwidth]{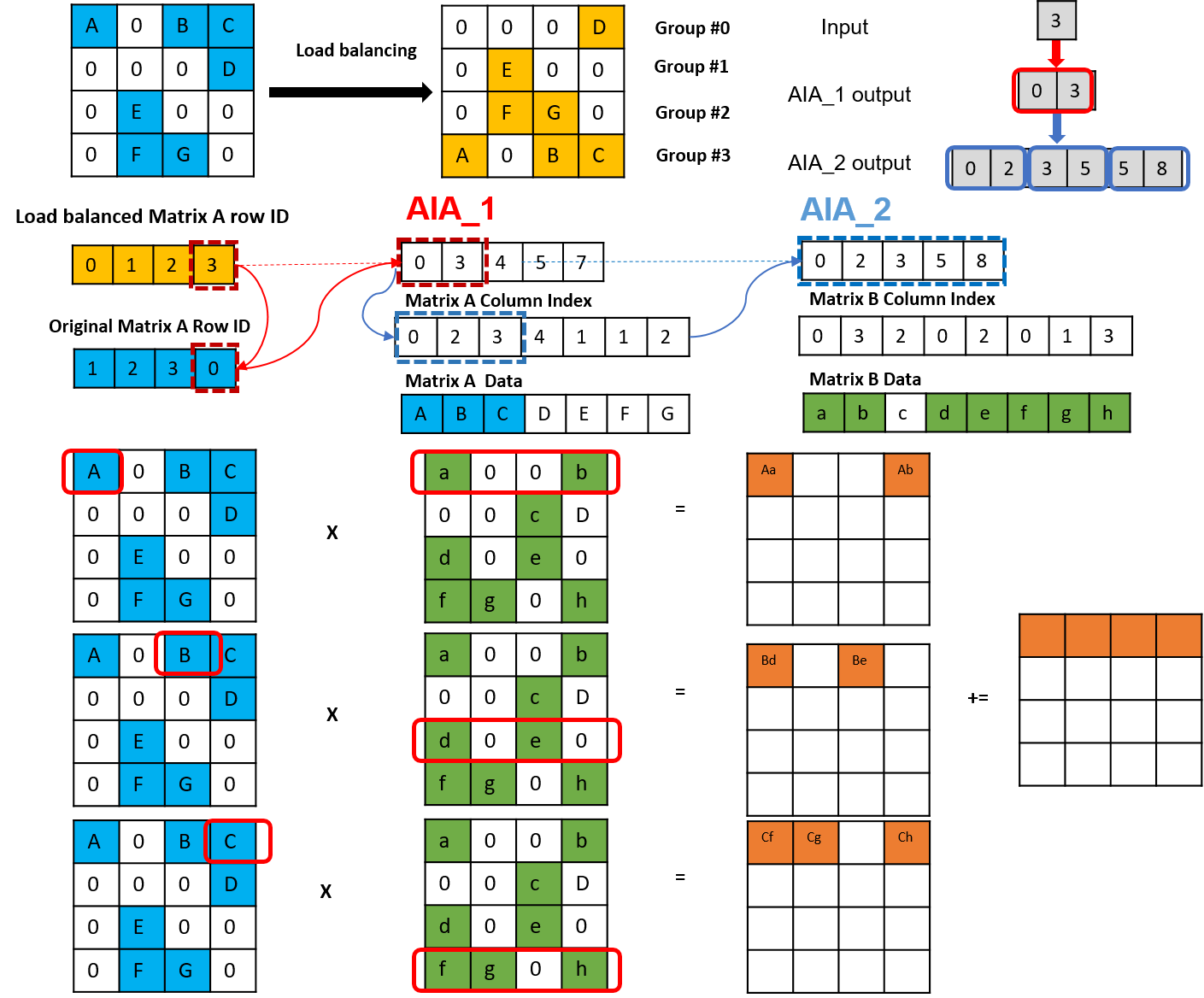}
\caption{AIA on SpGEMM: A Toy Example}
\label{fig:spgemm-AIA-range-2}
\end{figure}

The GPU kernel for SpGEMM is integrated using the AIA (ranged index $R=2$) approach. Each thread processes a row of matrix $\mathbf{A}$, using AIA to efficiently gather the required data from matrix $\mathbf{B}$. The AIA-range2 technique effectively addresses the fundamental challenge of irregular memory access patterns inherent in SpGEMM operations by transforming the two-level indirection problem into optimized sequential data streams. In SpGEMM, computing the result matrix $\mathbf{C}=\mathbf{A}\mathbf{B}$ requires accessing matrix $\mathbf{B}$'s rows corresponding to the non-zero column indices in each row of matrix $\mathbf{A}$, creating a complex indirect access pattern. As illustrated in Figure~\ref {fig:spgemm-AIA-range-2}, the AIA solution maps GPU block IDs to matrix rows through load-balanced mapping, where each group of rows has similar computational complexity. For block $i$ processing row $j$, the first AIA-range2 function computes $aia\_1[2i] = rpt_A[Map[i]]$ and $aia\_1[2i+1] = rpt_A[Map[i]+1]$ to define the range of non-zero elements in matrix $\mathbf{A}$, while the second AIA-range2 function $aia\_2[2j] =rpt_B[col_A[j]]$ and $aia\_2[2j+1] = rpt_B[col_A[j]+1]$ define the corresponding ranges in matrix $\mathbf{B}$. This approach enables the GPU cores to efficiently gather the required data from HBM in a more predictable pattern, significantly improving cache utilization and reducing memory latency. The grouping strategy enhances load balancing by ensuring that threads within the same block process rows with similar sparsity characteristics, while the ranged indirect access mechanism converts the traditionally irregular SpGEMM memory access patterns into optimized sequential streams that better match the GPU's memory hierarchy design.


This kernel structure allows for efficient parallelization of the SpGEMM operation while leveraging the AIA capabilities to handle irregular memory accesses. By integrating AIA functionality into the HBM-equipped GPU architecture, our solution provides a powerful platform for accelerating SpGEMM operations, addressing the key challenges of irregular memory access and load balancing in sparse matrix computations.

%% file: sections/algorithm.tex
\section{Benchmarks/Applications}
This section introduces the benchmarks and applications we evaluate to demonstrate the effectiveness of our AIA-accelerated SpGEMM implementation. We focus on key graph algorithms and sparse matrix operations that frequently occur in scientific computing and data analytics. Typical candidates include Markov Clustering (MCL), and Graph Contraction, and the widely used Graph Neural Network (GNN). In each application, we explain where and how SpGEMM can be applied.

\subsection{MCL}
The Markov Cluster (MCL) algorithm is an unsupervised graph clustering method that simulates stochastic flows on graphs to detect natural clusters. It operates on the principle that random walks on a graph will tend to get trapped within densely connected regions, corresponding to clusters. The algorithm begins by converting the input graph into a column stochastic matrix. It then iteratively applies two main operations: expansion and inflation. The expansion step, implemented as matrix multiplication or exponentiation, simulates random walks and allows flow to spread out. The inflation step, realized through the Hadamard power followed by normalization, strengthens intra-cluster connections while weakening inter-cluster connections. A crucial intermediate step, pruning, is applied to maintain sparsity by removing small entries and retaining only the top-k entries in each column. This process is repeated until convergence, typically when changes between successive iterations become negligible. The final matrix is interpreted to extract clusters, often represented as the connected components of the graph implied by the matrix. MCL's effectiveness in detecting clusters in various network types, coupled with its mathematical elegance, has made it a popular choice in bioinformatics and other fields dealing with complex network data.

\begin{algorithm}[ht]
\caption{Markov Cluster (MCL) Algorithm}
\label{alg:mcl_detailed}
\SetAlgoLined
\KwIn{Weighted network $G$, expansion parameter $e$, inflation parameter $r$, pruning threshold $\theta$, $k$} 
\KwOut{$clusters$}
$\text{AddSelfLoops}(G)$  \\
$A \gets \text{ColumnStochasticMatrix}(G)$\\
$\text{ColumnNormalize}(A)$\\

\While{change in successives iterations}{
    \tcp{Expansion: random walks from all vertices}
    $B \gets A^e$\\
    
    \tcp{Prune: Sparsify the matrix}
    $C \gets \text{Prune}(B, \theta, k)$\\
    \For{each column $j$ in $C$}{
        Remove entries $C_{ij} < \theta$\\
        Keep only top-$k$ entries in column $j$\\
    }
    
    \tcp{Inflation: strengthen intra-cluster connections}
    \For{each entry $C_{ij}$ in $C$}{
        $C_{ij} \gets (C_{ij})^r$\\
    }
    
    $A \gets \text{ColumnNormalize}(C)$\\
}

$clusters\gets \text{InterpretMatrix}(A)$\\
\Return $clusters$
\end{algorithm}

\subsection{Graph Contraction}
Graph contraction is a fundamental operation in graph theory and algorithm design that reduces the size of a graph by merging nodes with shared labels. This process is particularly useful in iterative graph algorithms where the problem can be solved on progressively smaller subgraphs. The algorithm efficiently implements contraction through matrix multiplication, utilizing a strategically constructed sparse matrix $S$. Given a graph $G$ and a set of node labels, the algorithm first determines the number of nodes $n$ and the maximum label value $m$. It then creates the sparse matrix $S$ of dimensions $m × n$, where each column corresponds to a node's original label, and each row represents a new label in the contracted graph. The matrix $S$ contains only ones, with their positions determined by the node labels. The contraction is then performed by the matrix multiplication $C = S × G × S^T$, where $G$ is the original graph's adjacency matrix and $S^T$ is the transpose of $S$. This operation effectively combines rows and columns of $G$ that share labels, resulting in a new, smaller adjacency matrix $C$ representing the contracted graph. The left multiplication by $S$ combines rows with the same label, while the right multiplication by $S^T$ combines columns, ensuring that edges between merged nodes are properly accounted for in the contracted graph.

\begin{algorithm}[ht]
\caption{Graph Contraction}
\label{alg:graph_contraction}
\SetAlgoLined
\KwIn{Graph $G$, node labels $labels$}
\KwOut{Contracted graph $C$}

$n \gets \text{length}(G)$\;
$m \gets \max(labels)$\;
$S \gets \text{sparse}(labels, 1:n, 1, m, n)$\;
$C \gets S \times G \times S^T$\;
\Return $C$
\end{algorithm}

\subsection{Graph Neural Network}
\begin{figure}[ht]
\centering
\includegraphics[width=0.48\textwidth]{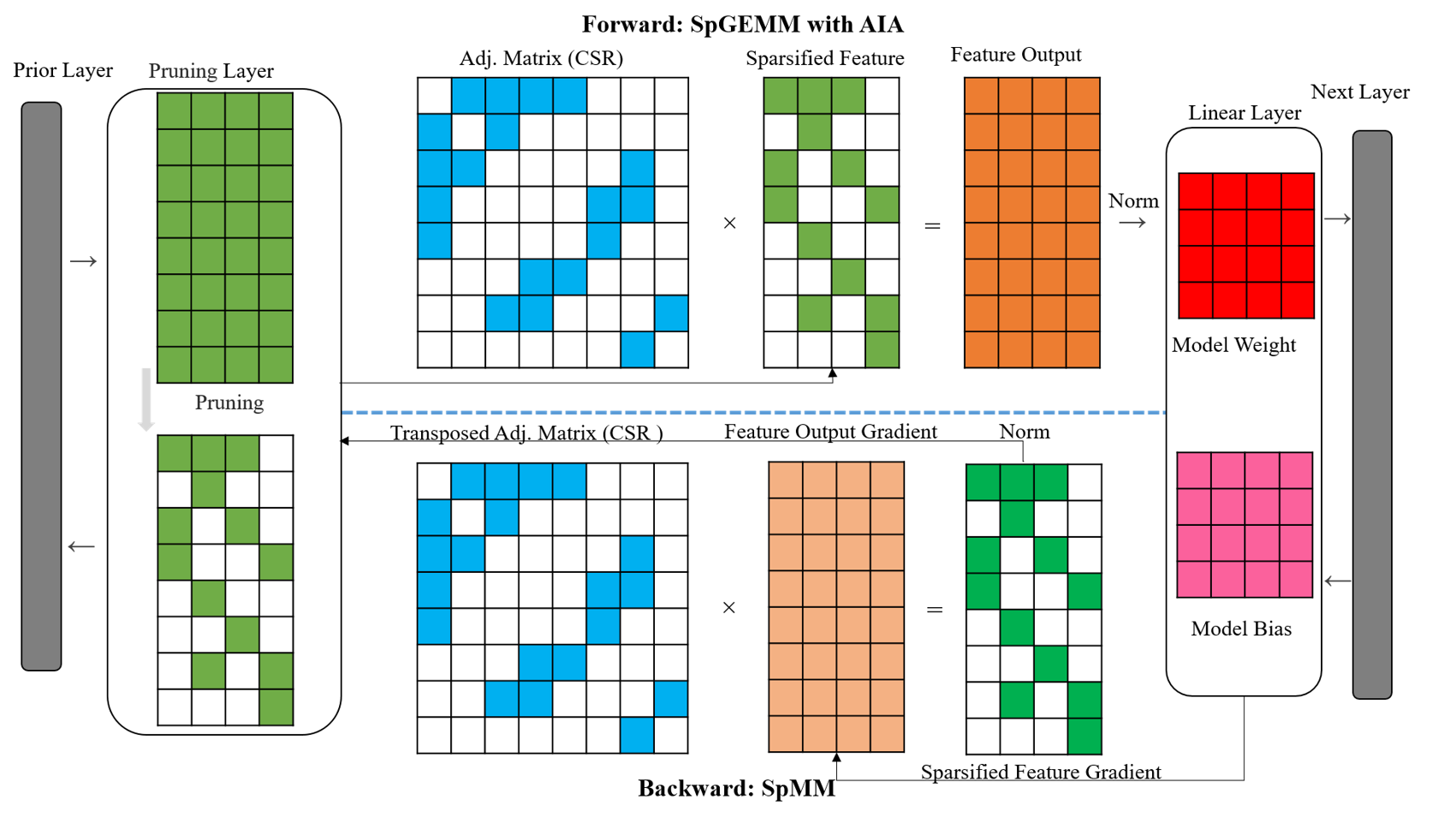}
\caption{GNN with Pruning Layer}
\label{fig:sparse_gnn}
\end{figure}

There are two promising approaches to accelerate Graph Neural Networks (GNNs) through sparse-sparse matrix multiplication (SpGEMM). One is to introduce pruning that sparsifies feature matrices, effectively transforming conventional SpMM operations into more efficient SpGEMM operations for full-batch training. The other is to formulate neighborhood sampling operations as Matrix-based Bulk sampling, where SpGEMM can be used for sampling in mini-batch training. 

GNN large graph full-batch training represents a challenging computational domain, particularly in forward propagation steps. Consider a graph $G = (V, E, A)$ containing $|V|$ nodes and $|E|$ edges, where the adjacency matrix $A \in \mathbb{R}^{|V| \times |V|}$ exhibits high sparsity. Traditional GNN forward propagation follows $X_l = A \cdot h(X_{l-1})$, where $X_l$ represents the node features at layer $l$, and $h(X_{l-1})$ denotes transformed features from the previous layer. These operations typically employ SpMM (Sparse-Dense Matrix Multiplication), becoming computationally intensive for large graphs.

In the forward pass of GNN using pruning techniques, the computation begins with features from the prior layer, which undergo processing through a pruning layer, say a top-K layer, that applies top-k selection-based sparsification to maintain only the most significant activations while eliminating lower-magnitude connections, effectively transforming conventional SpMM operations into more efficient SpGEMM operations, significantly reducing global memory traffic compared to traditional SpMM. The forward computation is mathematically reformulated as:

\begin{equation}
X_l = A \cdot \text{TopK}(X_{l-1}, k)W_l
\end{equation}

where $\text{TopK}(X_{l-1}, k)$ retains only the top-k values per sample or globally, creating a sparse representation:

\begin{equation}
\text{TopK}(X_{l-1}, k) = X_{l-1} \odot M_k
\end{equation}

with $M_k$ being a binary mask where $M_k[i,j] = 1$ if $X_{l-1}[i,j]$ is among the top-k values, and 0 otherwise.

The corresponding backward propagation also benefits from sparsity, expressed as:

\begin{equation}
\frac{\partial \mathcal{L}}{\partial X_{l-1}} = M_k \odot \left(A^T \cdot \frac{\partial \mathcal{L}}{\partial X_l}W_l^T\right)
\end{equation}

where the mask $M_k$ from the forward pass ensures gradients flow only through the top-k selected elements, naturally implementing gradient sparsity without additional computation. This winner-take-all gradient routing reinforces the importance of high-magnitude activations during training while maintaining computational efficiency through sparse operations.
For mini-batch training scenarios, the matrix-based bulk sampling approach expresses GNN neighborhood sampling as a series of SpGEMM operations. This framework formulates various sampling algorithms using the same matrix abstraction, as outlined in Algorithm~1. Each algorithm takes as input: (1) a sparse adjacency matrix $A$, (2) a sparse sampler-dependent matrix $Q^L$ that holds minibatch vertices, (3) batch size $b$, and (4) sampling parameter $s$. The output is a list of sampled adjacency matrices $A^0 \ldots A^{L-1}$ for a minibatch, where each $A^l$ is used in layer $l$'s aggregation step during forward propagation.

The algorithm follows a consistent three-step framework for each layer $l$ from $L$ down to 1: First, it computes probability distributions via SpGEMM as $P \leftarrow Q^l A$, where matrix $P$ holds one distribution per row. Second, it applies algorithm-specific normalization with $\text{NORM}(P)$. Third, it performs sampling with $Q^{l-1} \leftarrow \text{SAMPLE}(P, b, s)$ using Inverse Transform Sampling to select $s$ non-zeros per row of $P$. Finally, it extracts the sampled adjacency submatrix with $A^l \leftarrow \text{EXTRACT}(A, Q^l, Q^{l-1})$ by selecting rows from vertices in $Q^l$ and columns from vertices in $Q^{l-1}$. Different sampling algorithms implement specialized versions of the $\text{NORM}$ and $\text{EXTRACT}$ functions. For GraphSAGE, each row of $P$ represents the neighborhood of a batch vertex. For LADIES, each row of $P$ represents the distribution across the aggregated neighborhood of the entire batch. This uniform representation enables extending the approach to sample multiple minibatches in bulk, enhancing efficiency through communication-avoiding SpGEMM algorithms in distributed settings.

This matrix-based approach enables efficient processing of multiple minibatches while reducing communication overhead through communication-avoiding SpGEMM algorithms when distributed across devices.

%% file: sections/result.tex
\section{Experimental Results}
\textbf{Hardware Configuration.} All experiments were conducted on a Linux server with a 3.2GHz AMD EPYC 9555 64-Core CPU and 1.5 TB RAM. Our AIA solution is implemented on an NVIDIA H200 GPU. \\
\textbf{Methodology.} We compare our AIA accelerated SpGEMM against the original implementation without the AIA technique and the classic baseline (cuSPARSE library). Performance is evaluated based on execution time or FLOPS for matrix multiplication, where the latter is calculated as twice the number of intermediate products divided by execution time.

\begin{table*}[t]
\vspace{-2mm}
\scriptsize
\centering
\resizebox{0.96\textwidth}{!}{
\begin{tabular}{|c|c|c|c|c|c|c|}
\hline
Name & Row & Non-zero & NNZ/row & Max NNZ/row & Intermediate product of $A^2$ & NNZ of $A^2$ \\ \hline
RoadTX & 1,393,383 & 3,843,320 & 2.8 & 51 & 12099370 & 3843320 \\ \hline
p2p-Gnutella04 & 10,879 & 39,994 & 3.7 & 497 & 180230 & 39994 \\ \hline
amazon0601 & 403,394 & 3,387,388 & 8.4 & 100 & 32373599 & 16,258,436 \\ \hline
web-Google & 916,428 & 5,105,039 & 5.6 & 4334 & 60687836 & 29,710,164 \\ \hline
scircuit & 170,998 & 958,936 & 5.6 & 353 & 8,676,313 & 5,222,525 \\ \hline
cit-Patents & 3,774,768 & 16,518,948 & 4.4 & 770 & 82,152,992 & 68,848,721 \\ \hline
Economics & 206,500 & 1,273,389 & 6.2 & 44 & 7,556,897 & 6,704,899 \\ \hline
webbase-1M & 1,000,005 & 3,105,536 & 3.1 & 4700 & 69,524,195 & 51,111,996 \\ \hline
wb-edu & 9,845,725 & 57,156,537 & 5.8 & 3841 & 1,559,579,990 & 630,077,764 \\ \hline
cage15 & 5,154,859 & 99,199,551 & 19.2 & 47 & 2,078,631,615 & 929,023,247 \\ \hline
Wind Tunnel & 217,918 & 11,634,424 & 53.4 & 180 & 626,054,402 & 32,772,236 \\ \hline
Protein & 36,417 & 4,344,765 & 119.3 & 204 & 555,322,659 & 19,594,581 \\ \hline
\end{tabular}
}
\caption{Matrix Data}
\label{tab:matrix_data}
\vspace{-6mm}
\end{table*}
\subsection{Matrix Self-product Performance}
We selected 10 square matrices from the University of Florida Sparse Matrix Collection, commonly used for evaluating sparse matrix computations on GPUs. Table \ref{tab:matrix_data} summarizes the characteristics of these matrices, which vary significantly in size, sparsity, and non-zero element distribution.

\begin{figure}[ht]
\centering
\includegraphics[width=0.48\textwidth]{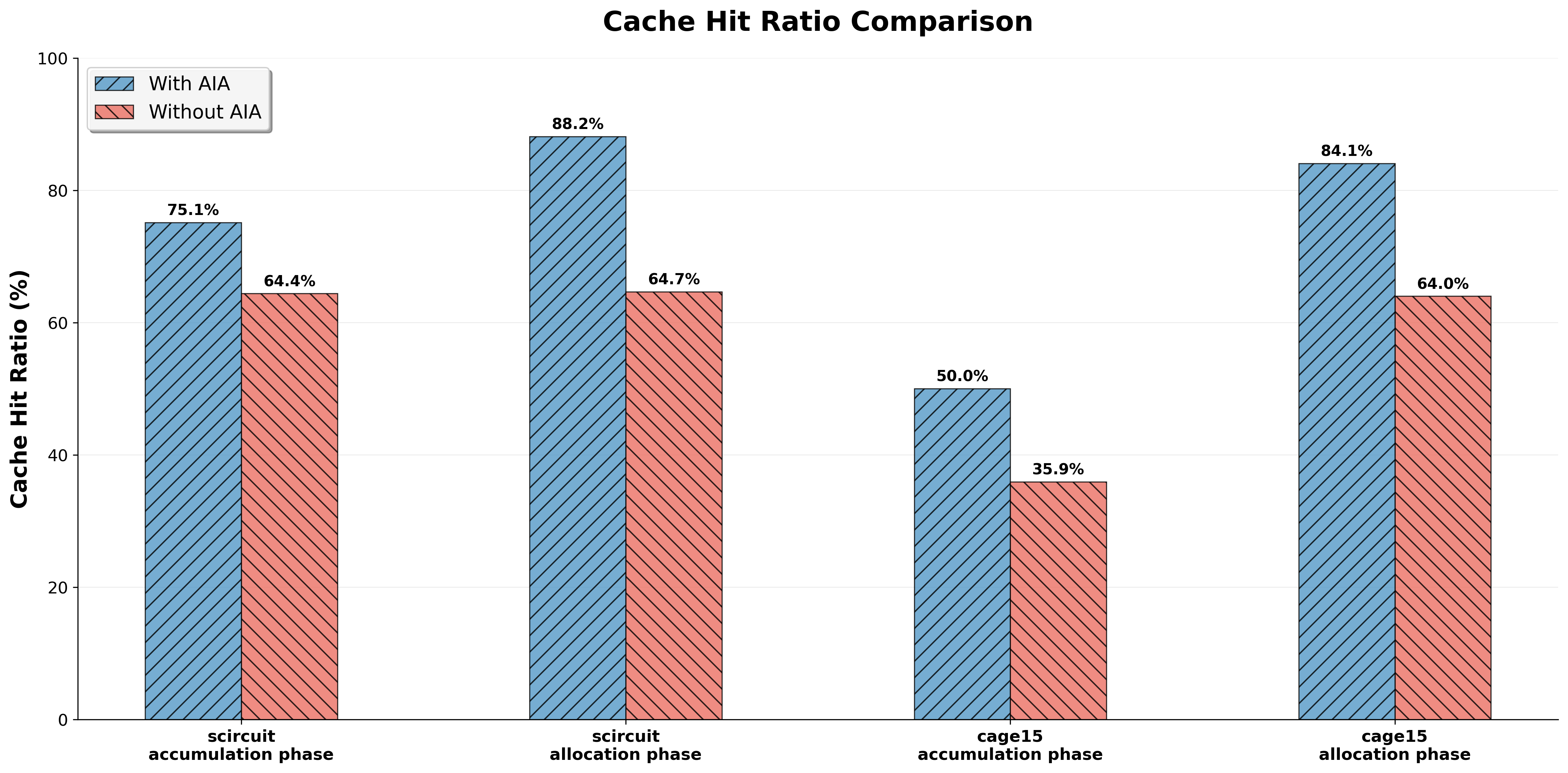}
\caption{L1 Cache Hit Ratio Comparison}
\label{fig:cache_hit_ratio}
\end{figure}

\textbf{Cache Hit Ratio:} Figure \ref{fig:cache_hit_ratio} shows the L1 cache hit ratio for the scircuit and cage15 datasets, comparing implementations with and without AIA across accumulation and allocation phases. AIA significantly improves cache utilization in both phases for both datasets. For scircuit, the hit ratio increases from 64.41\% to 75.14\% in the accumulation phase and from 64.66\% to 88.15\% in the allocation phase. Similarly, for cage15, improvements are observed from 35.94\% to 50.02\% in the accumulation phase and from 64.01\% to 84.10\% in the allocation phase. These results demonstrate AIA's effectiveness in optimizing memory access patterns, particularly in the allocation phase.

\begin{figure}[ht]
\centering
\includegraphics[width=0.48\textwidth]{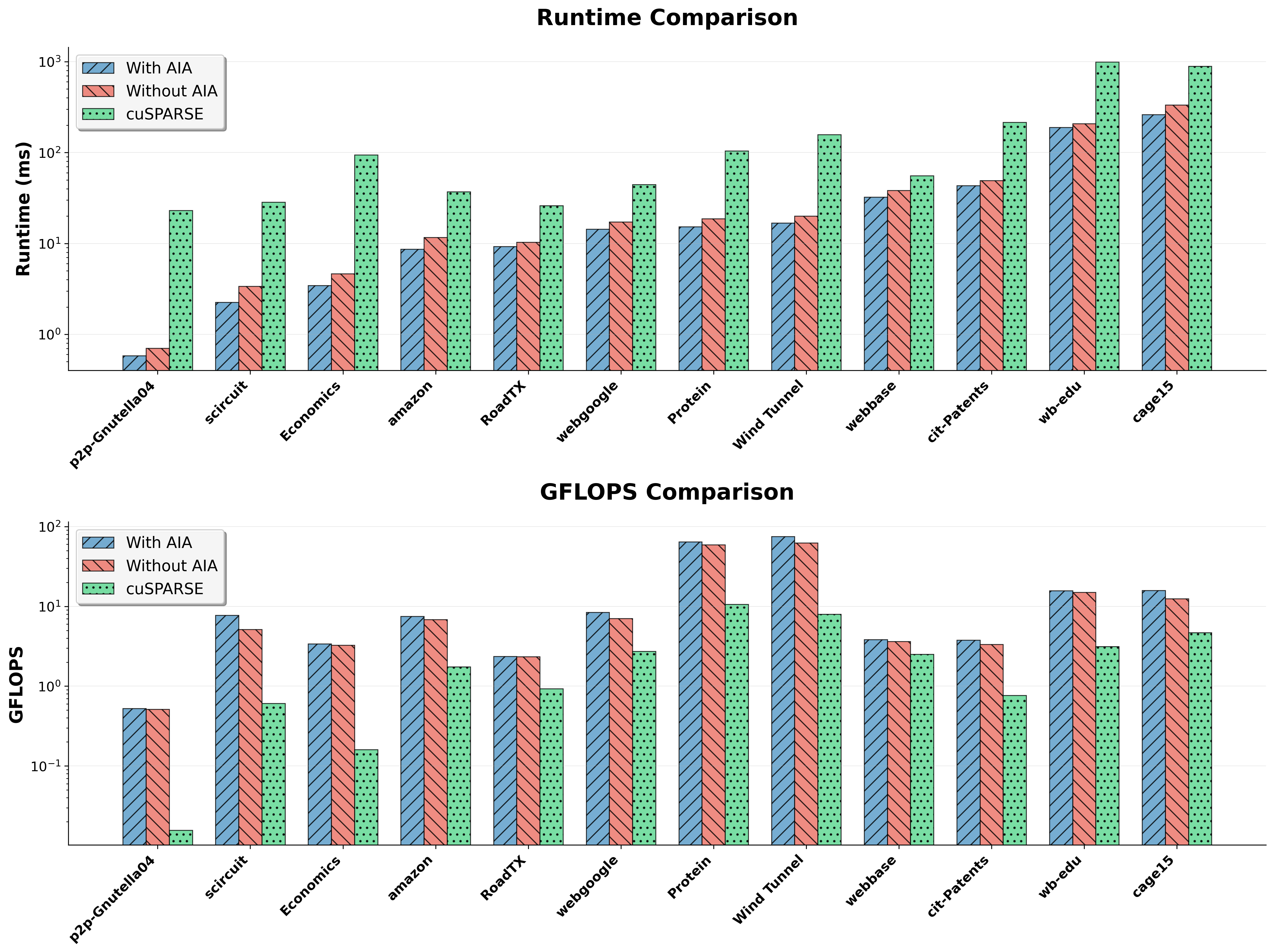}
\caption{Runtime and GFLOPS Comparison}
\label{fig:runtime_gflops}
\end{figure}

\textbf{Runtime and GFLOPS Analysis:} Figure \ref{fig:runtime_gflops} presents the runtime and GFLOPS performance comparison of matrix self-product operations. The AIA-enhanced implementation achieves an average 80.5\% runtime reduction compared to cuSPARSE, with execution times dropping from 240.48ms to 46.84ms on average. Individual improvements range from moderate gains on smaller matrices to dramatic speedups on large-scale problems—cage15 shows a 70.5\% reduction (262.5ms vs 888.4ms) while wb-edu demonstrates 81.0\% improvement (189.0ms vs 993.0ms). The GFLOPS comparison reveals a 6.87× average throughput improvement over cuSPARSE, with exceptional gains on scientific datasets like Wind Tunnel (9.4× speedup) and Protein (6.1× speedup). In runtime measurements, AIA achieves reductions ranging from 10-27\% compared to the software-only approach, with particularly notable improvements on large-scale matrices. These consistent performance advantages across diverse matrix structures, from road networks to web graphs, validate the effectiveness of our hardware acceleration in optimizing irregular memory access patterns inherent in SpGEMM operations.

\subsection{Graph Analytical Workload Performance}

\begin{figure}[ht]
\centering
\includegraphics[width=0.48\textwidth]{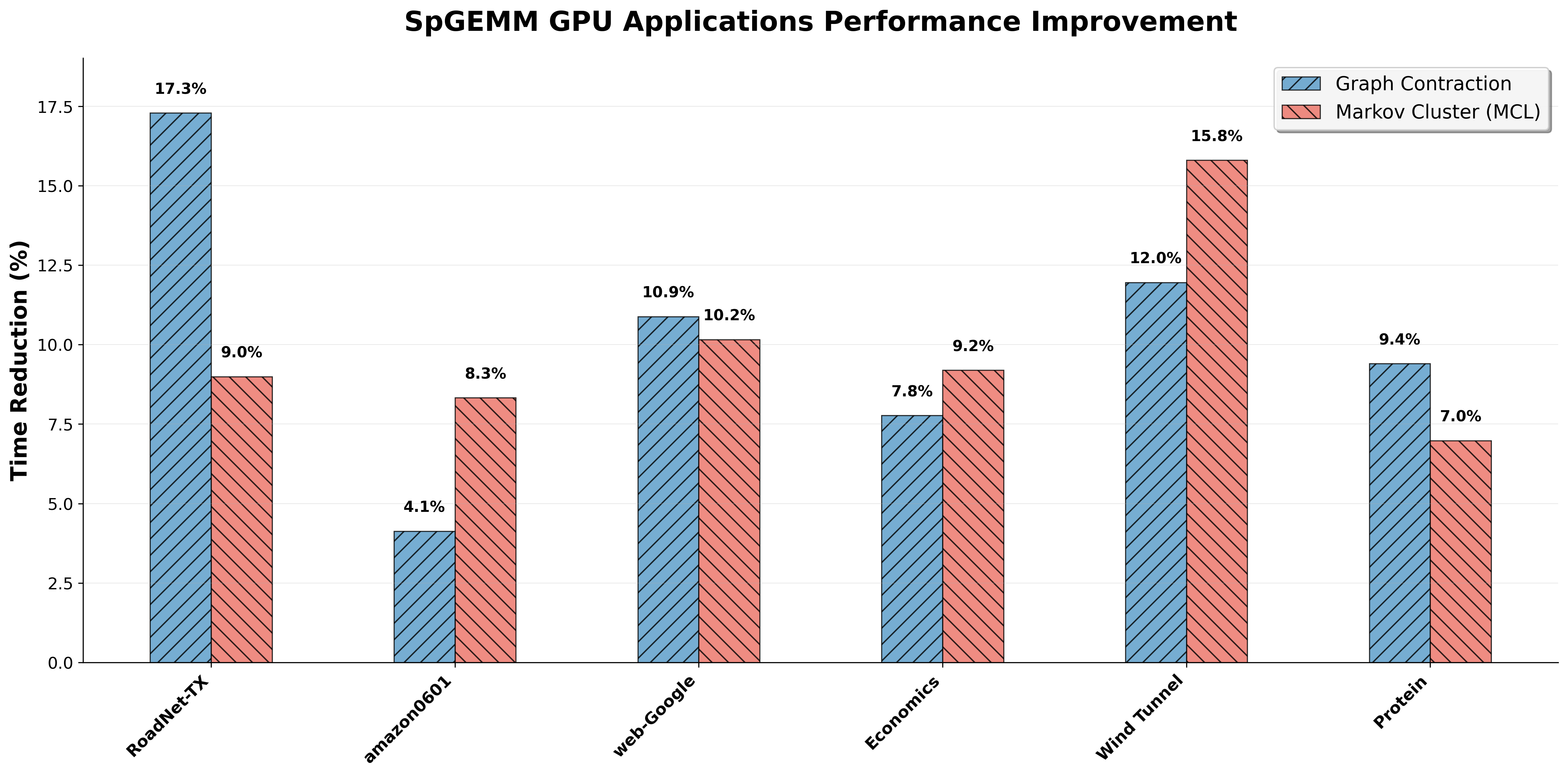}
\caption{SpGeMM Graph Applications Performance Improvement (AIA vs. Without AIA)}
\label{fig:runtime_gflops_noaia}
\end{figure}

\begin{figure}[ht]
\centering
\includegraphics[width=0.48\textwidth]{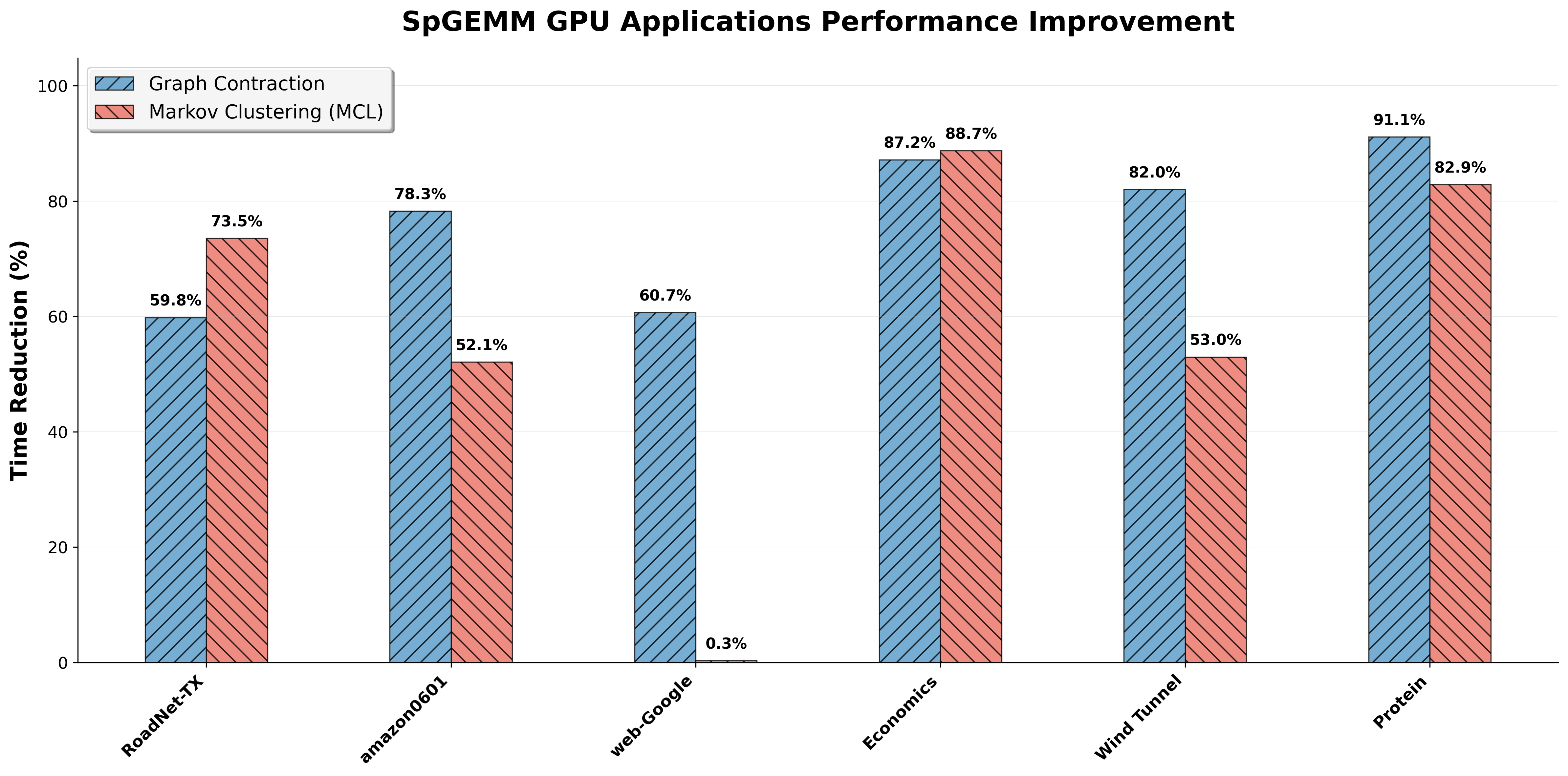}
\caption{SpGeMM Graph Applications Performance Improvement (AIA vs. cuSPARSE)}
\label{fig:runtime_gflops_cusparse}
\end{figure}

Figure~\ref{fig:runtime_gflops_noaia} and Figure~\ref{fig:runtime_gflops_cusparse} illustrate the performance improvements achieved by AIA over without-AIA and cuSPARSE for two graph applications: Graph Contraction and Markov Clustering (MCL). For Graph Contraction, AIA shows significant improvements ranging from 4.1\% to 17.3\%. The most notable improvements are observed with RoadNet-TX (17.3\%) and Wind Tunnel (12.0\%) datasets. Web-Google shows 8.9\% improvement, while Protein, Economics, and amazon0601 demonstrate 7.4\%, 5.8\%, and 4.1\% improvements, respectively. Markov Clustering exhibits strong performance gains across all datasets. Wind Tunnel dataset achieves the highest improvement at 13.8\%, followed by web-Google at 10.2\%. Other datasets show consistent gains: RoadNet-TX (9.0\%), amazon0601 (8.3\%), Economics (7.2\%), and Protein (5.0\%). MCL's improvement percentages, particularly for complex datasets like Wind Tunnel and web-Google, suggest that AIA is especially effective for iterative graph algorithms with multiple SpGEMM operations.

Comparing our AIA-enhanced SpGEMM implementation against cuSPARSE for Graph Contraction and Markov Clustering, the AIA approach achieves significant speedups, averaging 76.5\% for Graph Contraction and 58.4\% for MCL, with peak improvements reaching 91.1\% (Protein) and 88.7\% (Economics), respectively. These consistent performance gains across diverse matrix structures, from road networks to scientific datasets, demonstrate the effectiveness of our hardware acceleration in transforming irregular memory access patterns into efficient sequential streams, yielding an overall average speedup of 67.5\% over cuSPARSE.
\subsection{Graph Neural Network Training Performance}
We evaluate the effectiveness of our SpGEMM implementation on a full-batch Graph Neural Network training workload containing pruning layers across six benchmark datasets in Table \ref{tab:dataset_stats} and three GNN architectures (GCN, GIN, and GraphSAGE). Our experimental analysis demonstrates substantial performance improvements over baselines and reveals important scalability characteristics.
\begin{table}[h]
\centering
\caption{Dataset Characteristics for SpGEMM Performance Evaluation}
\label{tab:dataset_stats}
\resizebox{\columnwidth}{!}{%
\begin{tabular}{@{}lrrrrr@{}}
\toprule
\textbf{Dataset} & \textbf{Nodes} & \textbf{Edges} & \textbf{Avg. Degree} & \textbf{Density (\%)} & \textbf{Category} \\
\midrule
Flickr           & 89,250         & 989,006        & 22.16   & 0.0248 & Social \\
ogbn-proteins    & 132,534        & 79,122,504     & 1,193.92 & 0.9005 & Biological \\
ogbn-arxiv       & 169,343        & 1,335,586      & 15.77   & 0.0093 & Citation \\
Reddit           & 232,965        & 114,848,857    & 985.99  & 0.4232 & Social \\
Yelp             & 716,847        & 13,954,819     & 38.93   & 0.0054 & Social \\
ogbn-products    & 2,449,029      & 126,167,053    & 103.05  & 0.0042 & E-commerce \\
\bottomrule
\end{tabular}%
}
\vspace{2mm}
\end{table}

\begin{figure}[ht]
\centering
\includegraphics[width=0.48\textwidth]{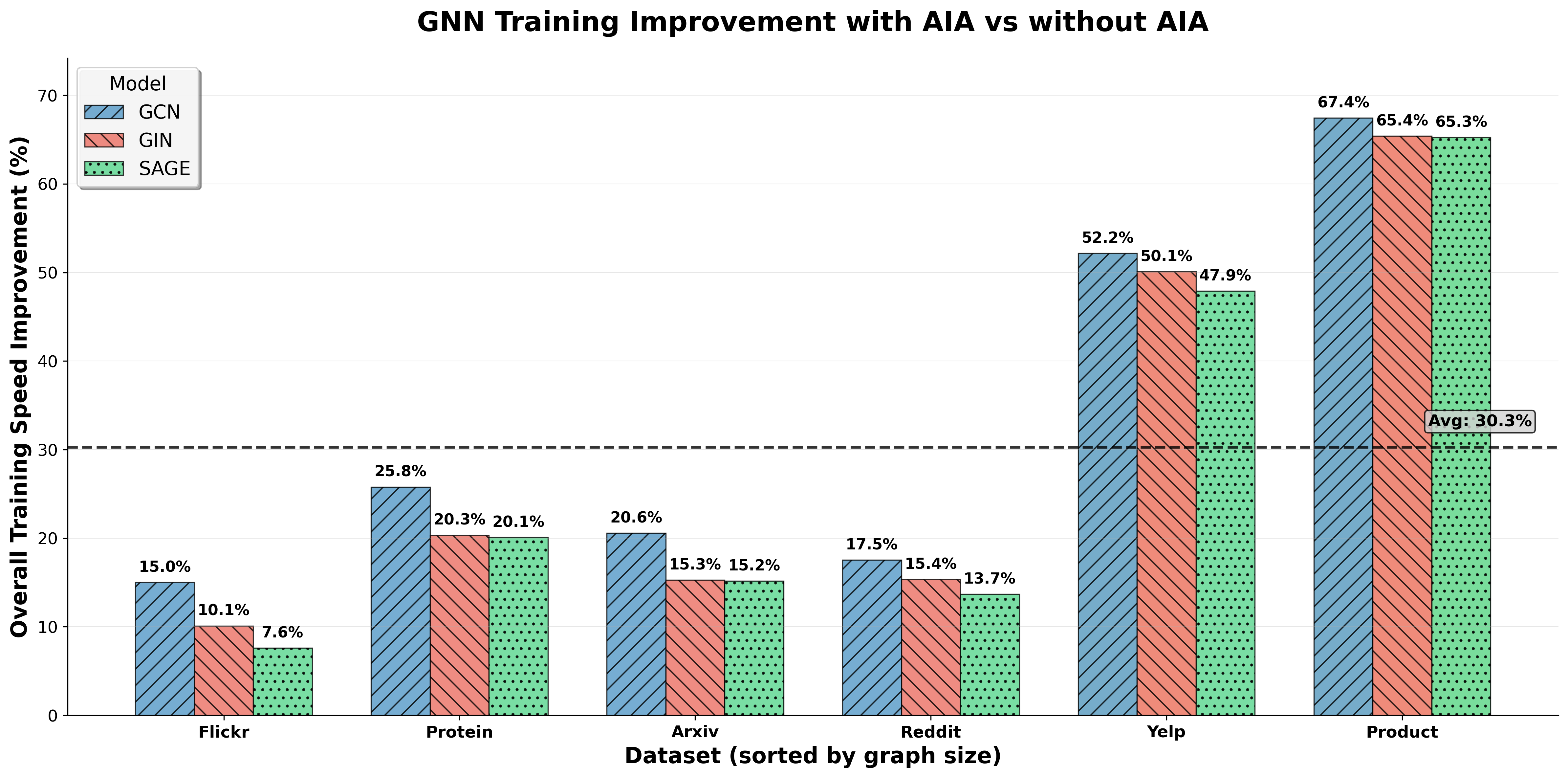}
\caption{Training time reduction ratio with AIA for prunned GNN (AIA vs. Without AIA)}
\label{fig:gnn_training_time_redu}
\vspace{-5pt}
\end{figure}

\noindent \textbf{Training Time Analysis with AIA.}
Figure~\ref{fig:gnn_training_time_redu} presents the training time improvements achieved through AIA acceleration. The results demonstrate consistent performance gains across all datasets and models, with an overall average improvement of 30.3\%. Notably, larger graphs exhibit more substantial improvements: Products (2.4M nodes) achieves 67.4\%, 65.4\%, and 65.3\% speedup for GCN, GIN, and SAGE respectively, while Yelp (717K nodes) shows 52.2\%, 50.1\%, and 47.9\% improvements.
The performance gains correlate positively with graph size, indicating superior scalability of our approach. Smaller graphs such as Flickr demonstrate modest improvements (15.0\%, 10.1\%, 7.6\%), while medium-sized graphs like Protein and Arxiv show intermediate gains of 20-25\%. This trend validates our hypothesis that AIA's benefits amplify with increasing memory access irregularity characteristic of larger graphs.

\begin{figure}[ht]
\centering
\includegraphics[width=0.48\textwidth]{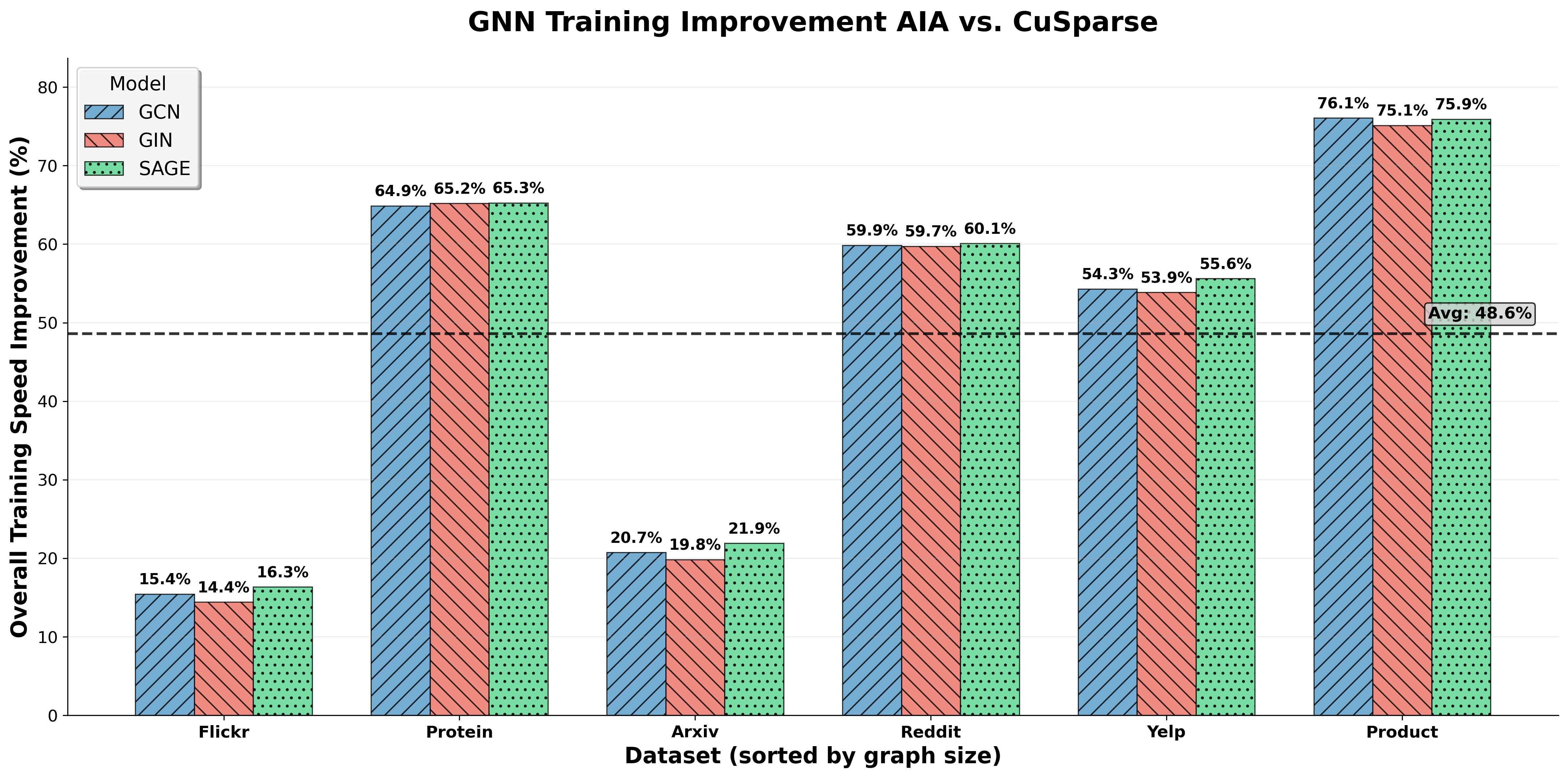}
\caption{Training time reduction ratio with AIA for prunned GNN (AIA vs. cuSPARSE)}
\label{fig:aia_time_redu_vs_cusparse}
\end{figure}

\noindent \textbf{AIA Performance Comparison with cuSPARSE.}
Figure~\ref{fig:aia_time_redu_vs_cusparse} quantifies the performance improvement of our AIA implementation compared to the cuSPARSE baseline. The results show an average improvement of 48.6\% across all configurations. The most significant gains are observed on the Products dataset (76.1\%, 75.1\%, 75.9\% for GCN, GIN, and SAGE respectively) and Protein dataset (64.9\%, 65.2\%, 65.3\%).
Reddit and Yelp datasets demonstrate consistent improvements in the 54-60\% range, while Arxiv shows moderate gains of approximately 20\%. The relatively uniform performance across different GNN architectures (standard deviation < 2\% for most datasets) indicates that our optimization is architecture-agnostic and broadly applicable to various GNN formulations.

\begin{figure}[ht]
\centering
\includegraphics[width=0.48\textwidth]{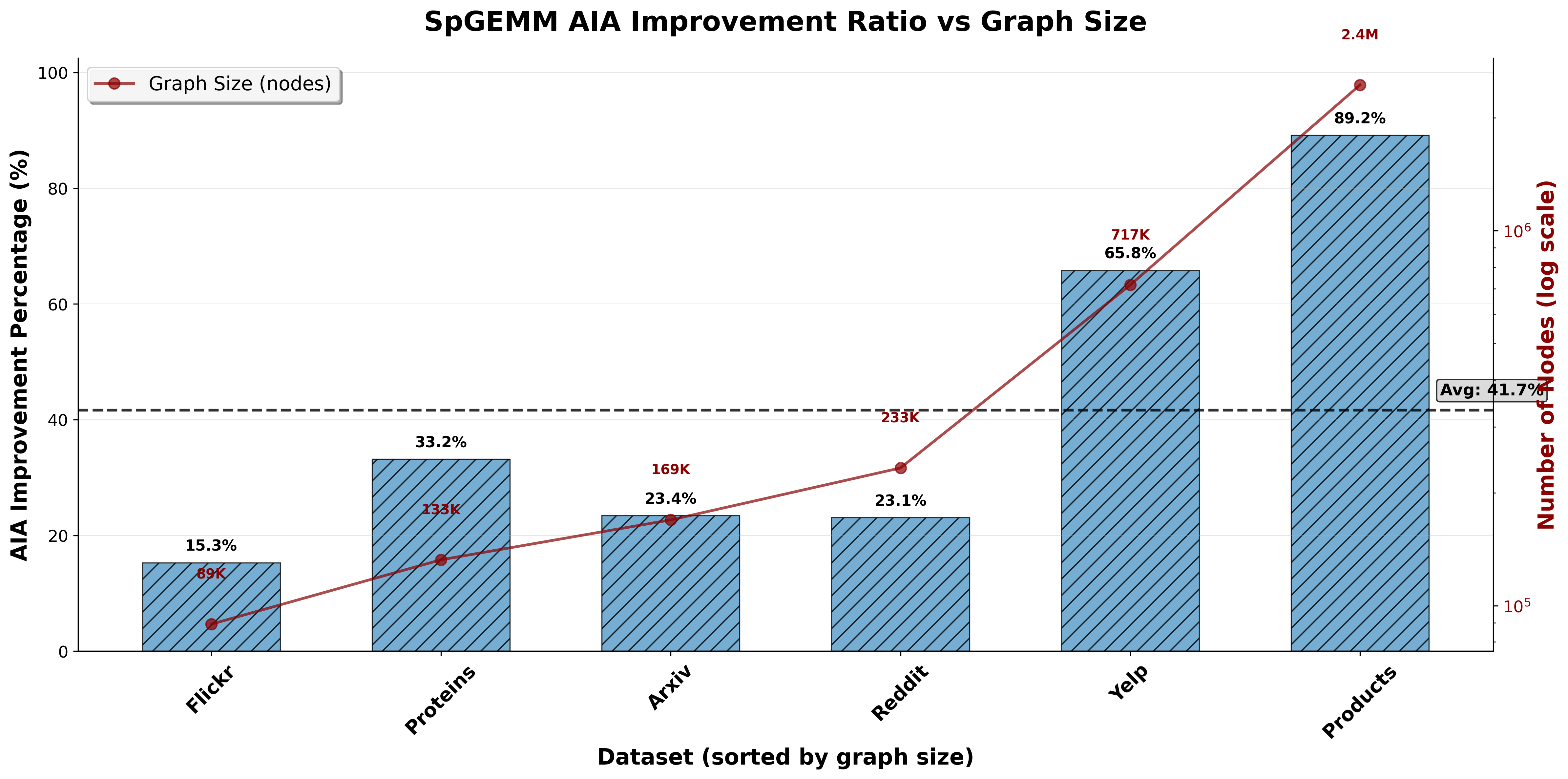}
\caption{SpGEMM AIA Time Reduction vs Graph Size (AIA vs. Without AIA)}
\label{fig:spgemm_time_redu}
\end{figure}

\noindent \textbf{Scalability Analysis.}
Figure~\ref{fig:spgemm_time_redu} illustrates the relationship between graph size and AIA improvement ratio on spgemm operation, revealing a strong positive correlation (Pearson's r = 0.94). The improvement percentage increases from 15.30\% for Flickr (89K nodes) to 89.16\% for Products (2.4M nodes), with an average improvement of 41.7\% across all datasets.
The superlinear scaling behavior observed for graphs exceeding 500K nodes can be attributed to two factors: (i) increased memory access irregularity in larger graphs amplifies the benefits of AIA's indirect access optimization, and (ii) the higher reuse distance in large graphs better utilizes AIA's prefetching capabilities. The exception observed for Reddit (233K nodes), showing 23.07\% improvement despite its size, suggests that graph topology and sparsity patterns also influence AIA effectiveness.

 AIA demonstrates superior scalability with graph size, achieving up to 89.16\% improvement for the largest tested dataset, confirming its effectiveness for production-scale GNN workloads. The consistent performance improvements across GCN, GIN, and GraphSAGE architectures (variance < 3\% within datasets) validate the generalizability of our approach across diverse GNN formulations. The combination of structured pruning with AIA acceleration yields multiplicative benefits, with total training time reductions averaging 30.3\% beyond pruning alone, demonstrating the complementary nature of algorithmic and architectural optimizations. These results establish our AIA-enhanced SpGEMM as an effective acceleration technique for GNN training, particularly valuable for large-scale graph applications where memory access patterns dominate computational bottlenecks.
\\\\
These improvements across different applications consistently demonstrate AIA's effectiveness in handling complex, application-specific workloads beyond basic SpGEMM operations. The results are particularly impressive for larger, more complex datasets and iterative algorithms, validating AIA's scalability and practical applicability in real-world graph processing scenarios.

%% file: sections/conclusion.tex
\section{Conclusion}
This paper presents the Acceleration of Indirect memory Access (AIA) , a novel Processing-Near-HBM solution for optimizing SpGEMM operations on GPUs. Our implementation achieves significant performance improvements across multiple evaluation metrics.
For basic SpGEMM operations, our hash-based implementation with AIA delivers 80.5\% runtime reduction and 6.87$\times$ throughput improvement compared to cuSPARSE. Cache utilization analysis reveals substantial improvements, with L1 cache hit ratios increasing from 64.41\% to 75.14\% in accumulation phases and from 64.66\% to 88.15\% in allocation phases. In practical graph applications, AIA shows up to 17.3\% runtime reduction over the non-AIA baseline. When compared to cuSPARSE, Graph Contraction achieves 76.5\% runtime reduction, and Markov Clustering achieves 58.4\% runtime reduction. For GNN training workloads with structured pruning, our approach delivers an average of 1.95$\times$ speedup, with large-scale datasets like Products (2.4M nodes) achieving up to 4.18$\times$ speedup while maintaining model accuracy.

These results establish AIA as an effective solution for accelerating sparse matrix operations on modern GPU architectures, particularly valuable for large-scale graph processing and iterative algorithms that rely heavily on SpGEMM operations.